\documentclass[11pt, letter]{article}
\pdfoutput=1
\usepackage[utf8]{inputenc}
        
\usepackage[square,numbers]{natbib}
\usepackage{authblk}

\usepackage{graphicx}              
\usepackage{amsmath}               
\usepackage{txfonts}
\usepackage{color}
\usepackage{hyperref}
\usepackage{subcaption}
\usepackage[frozencache,cachedir=.]{minted}

\newcommand{\ldbrac}[1]{\lvert#1\rangle}
\newcommand{\abs}[1]{\left\vert#1\right\vert}

\newcommand{\Romnum}[1]{\uppercase\expandafter{\romannumeral #1}}

\def\C{\ensuremath{\varmathbb{C}}}
\def\Z{\ensuremath{\varmathbb{Z}}}

\begin{document}
\title{Is Your Quantum Program Bug-Free?}

\author[1]{Andriy Miranskyy}

\author[1]{Lei Zhang}

\author[1]{Javad Doliskani}

\affil[1]{Department of Computer Science, Ryerson University \protect\\ Toronto, Canada}
\affil[]{{\{avm, leizhang, javad.doliskani\}@ryerson.ca}}

\date{}
\maketitle

\begin{abstract}
Quantum computers are becoming more mainstream. As more programmers are starting to look at writing quantum programs, they face an inevitable task of debugging their code. How should the programs for quantum computers be debugged?

In this paper, we discuss existing debugging tactics, used in developing programs for classic computers, and show which ones can be readily adopted. We also highlight quantum-computer-specific debugging issues and list novel techniques that are needed to address these issues. The practitioners can readily apply some of these tactics to their process of writing quantum programs, while researchers can learn about opportunities for future work.
\end{abstract}

\section{Introduction}\label{sec:intro}

Quantum Computers (QC) are specialized devices that will be able to solve some problems faster than Classic Computers (CC)~\cite{bernstein1997quantum, deutsch1985quantum}. This is known as a `quantum advantage'.

The QC field is still in its infancy: the largest machines built to date are of the order of tens of qubits~\cite{ibm_quantum,google2018}, which is not sufficient for commercially viable applications. However, the power of the QC increases. We do not know when the quantum advantage will be reached: predictions of experts vary from months ~\cite{Nevenlaw}, to 3-5 years~\cite{ibm2022}, to 7+ years~\cite{mosca2018cybersecurity}. This implies that QC may become practical within a decade. 

The programming languages for QC are mainly low-level, operating at the level of QC register (e.g., OpenQASM~\cite{cross2017open}). However, higher-level languages are being developed (e.g., Scaffold~\cite{JavadiAbhari2014ScaffCC}). 

It is argued~\cite{miranskyy2019testing} that for the foreseeable future, the QC will be used in a System of Systems, where the majority of software features will be implemented on a CC, while the features involving quantum algorithms will be `outsourced' to QC components. 

To enable usage of the QC, libraries with pre-packaged quantum algorithms start to appear. For example, Qiskit Aqua \cite{Qiskit} (an open source library written in Python) implements quantum algorithms for various domains, such as artificial intelligence, chemistry, and finance. Such a library enables a programmer to treat QC as a black-box and leverage quantum algorithms without having a deep understanding of the QC field.

Of course, the libraries themselves have to be developed by programmers with the expertise in the QC field. These programmers, inevitably, inject defects in their code (uniting CC and QC programming worlds). After that, the code has to be debugged. In this paper, we explore how existing debugging tactics can be applied to QC programs and which novel approaches have to be created.    

\section{Debugging Tactics}\label{sec:traditional}
Debugging is a process of removing an error, once this error has been exposed~\cite{pressman2014software}. While we hope that one day debugging will become an orderly and automated process (e.g., by automatically mapping bug reports to code where the defect resides, and then issuing a patch for this code~\cite{TufanoPWBP19}), currently it is an art more than a science~\cite{pressman2014software}. 

The high-level tactics~\cite[Chapter 8]{myers2011art} for debugging a software had not changed significantly over the last 40 years (when the first edition of the seminal work~\cite{myers2011art} was published), although integrated development environments and various automation tools have streamlined a lot of mundane tasks~\cite{zeller09debugging, MargineanBCH0MM19}. The three common tactics~\cite{myers2011art,pressman2014software} are backtracking, cause elimination, and brute force, discussed below.

\textit{Backtracking} debugging centers around examining the execution tree from the point of the error until a perpetrating code block is found. The analysis techniques for a code listing (such as code reviews and inspections) of a CC program can be readily applied to a QC program~\cite{miranskyy2019testing}. Thus, these tactics are transferable. Anecdotally, based on discussions with practitioners, code reviews and inspections are the most popular debugging techniques of quantum programs nowadays.

\textit{Cause elimination} debugging formulates a hypothesis (using inductive or deductive reasoning), specifying a root cause for a bug under study. Then, data are devised, and experiments are conducted to refute or prove this hypothesis. This approach can be applied to QC. Given the probabilistic nature of the QC programs~\cite{nielsen2010quantum, miranskyy2019testing}, we will have to execute the program multiple times to obtain a distribution of the results and assess the accuracy of the answer. Thus, we may be able to extend the techniques used for testing probabilistic programs running on CC, such as~\cite{DuttaLHM18, DuttaZHM19}, to the QC domain.

\textit{Brute force} debugging --- centred around the analysis of runtime traces, memory dumps, and output statements ---  focuses on runtime data analysis. Of the three tactics, this is the most common one~\cite{pressman2014software}. Some of the analyses of the runtime artifacts can be automated; however, a lot of the brute force debugging is still performed manually~\cite{pressman2014software}. Can we transfer these tactics? 

If we treat a QC program as a black-box, then the short answer is `yes'.  As mentioned in Section~\ref{sec:intro}, if a QC program will be used as part of a System of Systems, then we can trace the input (passed from a CC component to the QC component) and the output (from the QC component to the CC component). The input and output data can be recorded in a log, and these data can be compared against the expected values. 

But what if we would like to analyze a QC program at runtime using a white-box approach, e.g., to capture the execution trace of a QC program or perform interactive debugging of the code executed on the QC? In such a case, the short answer is `it depends'. Let us elaborate on this answer below.

\section{Debugging Quantum Programs}

A quantum program executed on a modern gate-based QC leverages a register of qubits for performing quantum operations and a register of classic bits for recording the measurements of qubits' states and conditionally applying quantum operators~\cite{cross2017open}. Thus, a typical QC program mixes traditional instructions (to alter the state of bits and apply conditional statements) and quantum instructions (to alter the state of qubits and to measure qubit value).

At any point of execution, the state of a CC is given by a vector of bits taking the values of $0$ and $1$. A register of $m$ bits can represent $2^m$ states. The state of a QC is, however, given by a vector of qubits and bits. 
A qubit is a two-state system and, thus, is an element of the space $\C^2$, where $\C$ is the set of complex numbers.  The quantum state of a qubit $\ldbrac{b}$, can be captured as a \textit{superposition} (linear combination) of the orthonormal basis states $\ldbrac{0}$ and $\ldbrac{1}$, where $\ldbrac{\cdot}$ denotes a vector in a vector space using the bra-ket notation~\cite{kaye2007introduction}. In the 
\textit{computational basis} (i.e. a combination of the $\ldbrac{0}$ and $\ldbrac{1}$ states), a qubit is written as a normalized vector $\ldbrac{b} = \alpha \ldbrac{0} + \beta \ldbrac{1}$, where $\abs{\alpha}^2 + \abs{\beta}^2 = 1$. 

A vector of qubits resides in a Hilbert space, which is an euclidean complex vector space (see~\cite{watrous2018theory} for details). An $n$-qubits state  is an element of $\C^{2^n}$.
A register of $n$-qubits is denoted by one of the equivalent notations $\ldbrac{b_1, \dots, b_n}$, 
or $\ldbrac{b_1} \ldbrac{b_2} \cdots \ldbrac{b_n}$, or $\ldbrac{b_1} \otimes \ldbrac{b_2} \otimes \cdots \otimes \ldbrac{b_n}$,
where $\otimes$ denotes the tensor product of two vectors.

As mentioned above, a general quantum program consists of blocks of code each containing classical and quantum instructions. 
Quantum operations can be divided into two kinds: unitary and non-unitary. Unitary operations are 
reversible and preserve the norm of the operands. Non-unitary operations are not reversible and have 
probabilistic implementations. 

The classical parts of a quantum program can be debugged using traditional methods. The quantum parts, 
however, can not be treated in the same way because of the properties of a QC~--- such as superposition, 
entanglement, and no-cloning~--- which are governed by the laws of quantum mechanics. The purpose of 
debugging a program is to present the user with human readable, i.e., classical, information about 
the runtime state of the system. Extracting classical information from a quantum state is done using 
measurement which is usually a non-unitary operation and results in collapse of the state, and hence 
an unintended behavior of the program. We shall describe, in the following, different scenarios in a 
QC to which classical debugging techniques cannot be applied, and discuss some potential solutions.

\subsection{Superposition}\label{sec:superposition}

\begin{figure}[h]
	\centering
	\begin{subfigure}[b]{0.5\columnwidth}
		\centering
		\includegraphics[width = \textwidth]{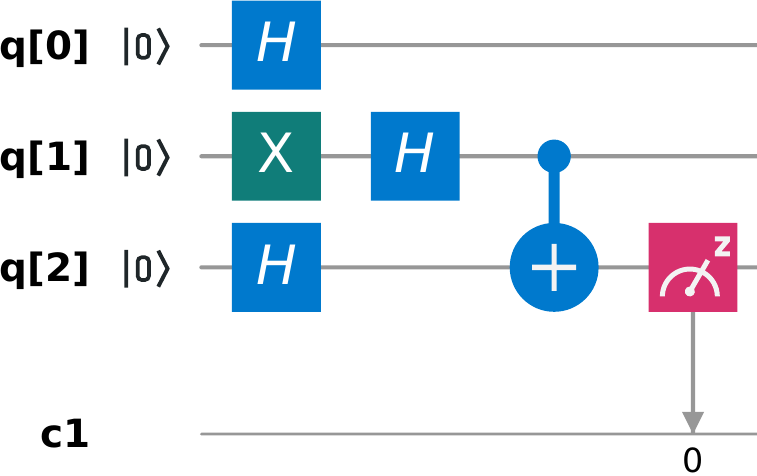}
		\caption{Circuit}
		\label{fig:spp-circ}
	\end{subfigure}
	\hfill
	\begin{subfigure}[b]{0.35\columnwidth}
		\centering
		\begin{minted}[fontsize = \footnotesize, numbersep = 2mm, linenos = true, autogobble]{cpp}
            OPENQASM 2.0;
            include "qelib1.inc";
            
            qreg q[3];
            creg c[1];
            
            x q[1];
            h q[0];
            h q[1];
            h q[2];
            cx q[1],q[2];
            measure q[2] -> c[0];
        \end{minted}
		\caption{Assembly code}
		\label{fig:spp-code}
	\end{subfigure}
	\caption{Example of measuring a superposition. In OpenQASM, NOT is denoted by $x$, Hadamard --- by $h$, and CNOT --- by $cx$. }
	\label{fig:spp}
\end{figure}

Let $\ldbrac{\psi}$ be the state of an $n$-qubit register. We can uniquely write $\ldbrac{\psi}$, in 
the computational basis, as
\[ \ldbrac{\psi} = \sum_{x \in \{ 0, 1 \}^n} \alpha_x \ldbrac{x}, \]
where $\alpha_x \in \C$ and $\sum_{x \in \{ 0, 1 \}^n} \abs{\alpha_x}^2 = 1$. We say that 
$\ldbrac{\psi}$ is a superposition of the basis states $\{ \ldbrac{x} \}_{x \in \{ 0, 1 \}^n}$. By 
the measurement postulate of quantum mechanics, measuring the state $\ldbrac{\psi}$ in the 
computational basis results in an outcome $x \in \{ 0, 1 \}^n$ with probability $\abs{\alpha_x}^2$, 
and the state of the system after the measurement is $\ldbrac{x}$. For example, consider the initial state
$\ldbrac{010}$ (which is set by applying NOT to qubit 2), and perform the following steps: first apply a Hadamard transform\footnote{The Hadamard transform on one qubit is given by $\ldbrac{b} \mapsto \frac{1}{\sqrt{2}}(\ldbrac{0} + (-1)^b \ldbrac{1})$. A description of transforms and associated quantum gates are given in~\cite{kaye2007introduction}.} to each qubit (creating superposition), then
a controlled-not (CNOT)\footnote{The CNOT transform is defined on two qubits and is given by $\ldbrac{a, b} \mapsto \ldbrac{a, a \oplus b}$, where $\oplus$ is
xor. Here, $\ldbrac{a}, \ldbrac{b}$ are called the control and input qubits, respectively.} to qubits 2 and 3, and finally measure qubit 3. If the measured qubit is $0$ 
(which happens with probability $1/2$), then the state collapses to $\frac{1}{2}(\ldbrac{00} - \ldbrac{01} 
+ \ldbrac{10} - \ldbrac{11})$. An implementation of this example in OpenQASM 2.0 is shown in Figure \ref{fig:spp}.

A natural feature of a debugger for quantum programs would be to check if the state 
of a variable is in superposition. There are the following two scenarios: when the input state is unknown (e.g., when it is generated as an output of another quantum program) and when the input state is known. Let us elaborate on these two cases.

\subsubsection{Unknown input state.}\label{sec:unknown-inp}
If the input to the program is an unknown state $\ldbrac{\psi}$, then there is no known general 
algorithm that can efficiently decide if $\ldbrac{\psi}$ is in a superposition. 
Not much can be done here in terms of a general method for debugging, 
different approaches should be considered for different problems.

For example, in the hidden subgroup problem \cite[Chapter 7]{kaye2007introduction}, if the group is abelian, then it can be efficiently 
decided if the coset state of a subgroup is in superposition. For non-abelian groups, however, the same problem is often hard. For example, 
the best known algorithm for the following problem has subexponential runtime \cite{kuperberg2005subexponential}: let $N$ be a positive 
integer, and let $\Z_N$ be the group of integers mod $N$. For a random unknown $x \in \Z_N$ and fixed unknown 
$d \in \Z_N$, decide whether a given state is of the form $\ldbrac{x}$ or $\frac{1}{\sqrt{2}}(\ldbrac{x} + \ldbrac{x + d})$.

\subsubsection{Known input state.}\label{sec:known-inp}
If a state is the result of applying a unitary operation to a known initial state, i.e., 
$\ldbrac{\psi} = U \ldbrac{x}$ where $x$ is known, then $\ldbrac{\psi}$ can be regenerated by the 
debugger. For example, the state 
\[ \ldbrac{\psi} = \frac{1}{\sqrt{2^n}} \sum_{x \in \{ 0, 1 \}^n} (-1)^{h(x)} \ldbrac{x}, \]
where $h(x)$ is the Hamming weight\footnote{That is, the number of non-zero bits in $x$.} of $x$, can be generated by applying the Hadamard transform to the $n$-qubit register $\ldbrac{11 \dots 1}$. In such cases, there are various methods (depending on 
the problem) to characterize the state $\ldbrac{\psi}$. Often, one relies on \textit{quantum state 
tomography}, which is the process of reconstructing a quantum state through a series of measurements \cite{d2003quantum, cramer2010efficient}.

\subsection{Entanglement}

In a QC, a set of memory cells or registers is said to be in an entangled state if it is impossible 
to classically specify the correlations among them. More precisely, let $X_1, \dots, X_n$ be the 
state spaces of a set of quantum systems that represent $n$ registers. The state space of the 
composite of these systems, that represents an array, is given by the tensor product $X = X_1 
\otimes \cdots \otimes X_n$. A state $\ldbrac{\psi} \in X$ that can be written in the form 
$\ldbrac{\psi} = \ldbrac{\psi_1} \otimes \cdots \otimes \ldbrac{\psi_n}$, where $\ldbrac{\psi_j} \in 
X_j$ for $j = 1, \dots, n$, is called separable. A state that is not separable is called entangled. 
When debugging a program that operates on an entangled state, the following problems can be 
considered.

\subsubsection{Checking for separability.}\label{sec:separability}
Given a state $\ldbrac{\psi} \in X$, deciding whether $\ldbrac{\psi}$ is separable is an NP-hard 
problem \cite{gharibian2010strong, gurvits2003classical}. This is called the \textit{separability 
problem} in quantum information theory, see \cite[Chapter 6]{watrous2018theory} for details. There are a variety of methods (see~\cite{leinaas2006geometrical,guhne2009entanglement}) for separability/entanglement detection 
that can be implemented in practice, specially for lower dimensions. For example, if the debugger can generate 
several copies of $\ldbrac{\psi}$, then one way to detect the nonlinear properties of 
$\ldbrac{\psi}$ is via direct measurement. For the sake of brevity, we do not provide technical details here; see \cite{leinaas2006geometrical} for a numerical method 
for examining separability and \cite{guhne2009entanglement} for other interesting methods and their 
implementations.

\subsubsection{Extracting classical information.}\label{sec:classical}
Measuring a subsystem of a larger composite system that is in an entangle state will likely alter 
other subsystems. This prevents a debugger from presenting any classical information about a 
variable (whose state is entangled) to the user without disturbing its state. 

For example, consider the entangled state $\frac{1}{\sqrt{2}}(\ldbrac{00} + \ldbrac{11})$ of two qubits.
Measuring any of the two qubits alters the result of the subsequent measurement on the other qubit.
More precisely, if the first qubit is measured, then state collapses to $\ldbrac{00}$ or $\ldbrac{11}$ with probability $|1/\sqrt{2}|^2 =
1/2$; the outcome of measuring the second qubit is always $0$ if the resulting state is $\ldbrac{00}$, and it is always $1$ if the resulting state is $\ldbrac{11}$. Such a state is called maximally entangled. 

\begin{figure}[h]
	\centering
	\begin{subfigure}[b]{0.5\columnwidth}
		\centering
		\includegraphics[width = \textwidth]{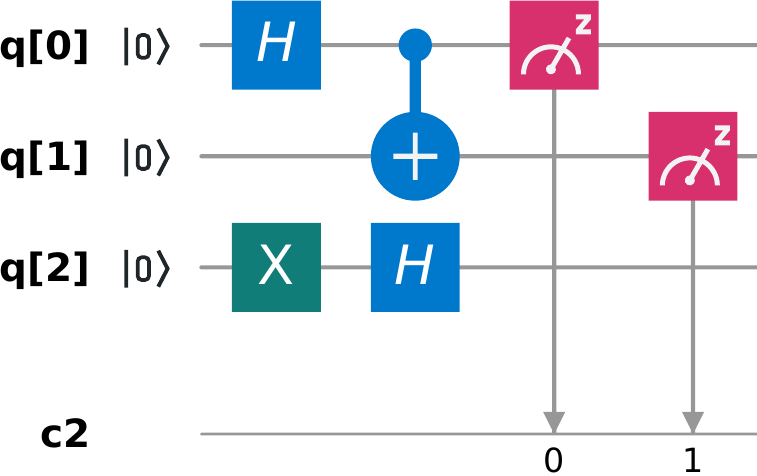}
		\caption{Circuit}
		\label{fig:entg-circ}
	\end{subfigure}
	\hfill
	\begin{subfigure}[b]{0.35\columnwidth}
		\centering
		\begin{minted}[fontsize = \footnotesize, numbersep = 2mm, linenos = true, autogobble]{cpp}
            OPENQASM 2.0;
            include "qelib1.inc";
            
            qreg q[3];
            creg c[2];
            
            x q[2];
            h q[0];
            cx q[0], q[1];
            h q[2];
            measure q[0] -> c[0];
            measure q[1] -> c[1];
        \end{minted}
		\caption{Assembly code}
		\label{fig:entg-code}
	\end{subfigure}
	\caption{An implementation of state \eqref{equ:gen-clas}.}
	\label{fig:entg-sep}
\end{figure}

A composite system, however, often has subsystems that are not entangled with any other subsystem. In 
this case, we can measure that subsystem without disturbing the whole state. For example, in the 
3-qubit register
\begin{equation}
\label{equ:sep-subsys}
	\frac{1}{2} (\ldbrac{000} - \ldbrac{001} + \ldbrac{110} - \ldbrac{111})
\end{equation}
the last qubit is not entangled with the first two while the first and the second qubits are 
entangled, see \eqref{equ:gen-clas}. The algorithms for separability detection (discussed in Section~\ref{sec:separability})  could be used to identify separable subsystems. 
Things would be much simpler if the debugger could somehow estimate a given state with a state that 
is generated by applying some operation to a basis state, i.e., classical information. For example, 
the state in \eqref{equ:sep-subsys} can be generated as
\begin{equation}
\label{equ:gen-clas}
\begin{split}
    (\operatorname{CNOT} \otimes H_2)&(H_2 \otimes I_4) \ldbrac{001} = \\
    & = \frac{1}{\sqrt{2}}(\ldbrac{00} + 
    \ldbrac{11}) \otimes \frac{1}{\sqrt{2}}(\ldbrac{0} - \ldbrac{1}),
\end{split}
\end{equation}
where $I_4$ and $H_2$ are the identity and Hadamard gates, respectively. Therefore, the state in \eqref{equ:sep-subsys} can be described by the debugger using the classical information 
$\ldbrac{001}$ and the names of the above operators. An implementation of the sequence of operations in \eqref{equ:gen-clas} is 
shown in Figure \ref{fig:entg-sep}.

\subsection{No-cloning}\label{sec:no-cloning}

The most general method of obtaining information about a variable without disturbing its state is 
to make a copy of the variable and work on the copy. In the classical setting, this is often 
straightforward. In the quantum setting, however, the situation is much more complicated. In fact, 
it is impossible to make a copy of a given general unknown quantum state. More precisely, given an 
unknown state $\ldbrac{\psi}$ and an arbitrary state $\ldbrac{\phi}$, it can be shown~\cite[Theorem 10.4.1]{kaye2007introduction} that 
there is no unitary operator $U$ that can perform the following:
\[ \ldbrac{\psi} \otimes \ldbrac{\phi} \overset{U}{\longmapsto} \ldbrac{\psi} \otimes 
\ldbrac{\psi}. \] 
In many practical scenarios, however, a debugger will only need to make an \textit{approximate copy} 
of a state; a state that is ``close enough'' to the given state but provides useful debugging 
information. For example, for a state $\ldbrac{\psi}$ that encodes a probability distribution \cite{grover2002creating}, such as
the Gaussian distribution, an approximate clone would provide valuable information about the distribution.
The possibility of approximate cloning was first discussed in \cite{buvzek1996quantum}. 
Much research has been done on different cloning methods each optimizing particular aspects of a 
cloner that are desired for different situations, see \cite{scarani2005quantum} for a survey.

\subsection{Discussion}
In Sections~\ref{sec:superposition}--\ref{sec:no-cloning}, we discussed various issues preventing the application of the classic debugging techniques and identified some potential solutions. 

As discussed in~\cite{miranskyy2019testing}, if the input size and the amount of required qubits is small, we can run a quantum program in a simulator (running on a CC). However, the increase of the input size and the qubit register length may force us to run the program on a QC. 

If we can generate multiple approximate copies of the state \cite{buvzek1996quantum}, then we can produce an empirical distribution of the qubit state and compare it against the expected distribution, to detect problems in the code. The generation of the multiple approximate copies can be readily implemented for moderate inputs sizes using universal cloning methods~\cite{werner1998optimal, buvzek1998universal, fan2001quantum}.  More efficient cloning can be achieved using state-dependent (i.e. non-universal) cloning methods~\cite{niu1999two, scarani2005quantum}. This would address issues related to superposition with known input state (discussed in Section~\ref{sec:known-inp}), extraction of classical information (discussed in Section~\ref{sec:classical}), and no-cloning (discussed in Section~\ref{sec:no-cloning}). A compiler can automatically generate the code for the approximate copying (akin to compilers for CC that can instrument the code to add debugging information), translating higher-level language into quantum assembly~\cite{huang2019statistical}. The same principle of multiple approximate copies can be used to generate runtime assertions~\cite{zhou2019quantum}.

For the case of unknown input states, discussed in Section~\ref{sec:unknown-inp}, no general solution exists and will require a programmer to make decisions on a case-by-case basis.

Finally, separability checking, discussed in Section~\ref{sec:separability}, demands the implementation of numerical methods that will require changes to the QC and, hopefully, will be implemented in the future. 

\section{Conclusions}

QC field is rapidly evolving, and the Software Engineering (SE) community should start bringing SE practices into the QC world.
In this paper, we focus on the analysis of debugging tactics, highlighting classic ones that are readily applicable and showing that new ones have to be created.
We believe that this work would be of interest to practitioners, creating quantum programs, as well as researchers, developing the next generations of tooling for QC.

\bibliography{references}

\end{document}